\documentclass[namedreferences,hyperref,optionalrh]{spr-sola}
\usepackage{graphicx}        
\usepackage[dvipsnames]{xcolor}           

\newcommand{\Rs}{R$_\odot$}
\newcommand{\degr}{$^\circ$}

\newcommand{\aj}{{\it Astron. J.}} 
\newcommand{\apj}{{\it Astrophys. J.}}
\newcommand{\apjl}{{\it Astrophys. J. Lett.}}

\newcommand{\pasp}{{\it Publ. Astron. Soc. Pac.}}

\newcommand{\solphys}{{\it Solar Physics}}
 
\newcommand{\ssr}{{\it Space Sci. Rev.}} 
\chardef\us=`\_

\begin{document}

\begin{frontmatter}
\title{Observations of the Polarized Solar Corona during the Annular Eclipse of October 14, 2023}

\author[addressref={aff1},corref]{\inits{D.B.}\fnm{Daniel~B.}~\snm{Seaton}\orcid{0000-0002-0494-2025}}
\author[addressref=aff1]{\inits{A.}\fnm{Amir}~\snm{Caspi}\orcid{0000-0001-8702-8273}}
\author[addressref={aff2,aff3}]{\inits{N.}\fnm{Nathalia}~\snm{Alzate}\orcid{0000-0001-5207-9628}}
\author[addressref=aff4]{\inits{S.}\fnm{Sarah J.}~\snm{Davis}\orcid{0009-0008-4901-0601}}
\author{\inits{A.R.}\fnm{Alec~R.}~\snm{DeForest}}
\author[addressref=aff1]{\inits{C.E.}\fnm{Craig~E.}~\snm{DeForest}\orcid{0000-0002-7164-2786}}
\author[addressref=aff1]
{\inits{N.F.}\fnm{Nicholas~F.}~\snm{Erickson}\orcid{0000-0001-6028-1703}}
\author[addressref=aff1]{\inits{S.A.}\fnm{Sarah~A.}~\snm{Kovac}\orcid{0000-0003-1714-5970}}
\author[addressref=aff1]{\inits{R.}\fnm{Ritesh}~\snm{Patel}\orcid{0000-0001-8504-2725}}
\author[addressref=aff1]
{\inits{S.N.}\fnm{Steven~N.}~\snm{Osterman}\orcid{0000-0000-0000-0000}}
\author[addressref=aff1]{\inits{A.}\fnm{Anna}~\snm{Tosolini}\orcid{0009-0000-3510-8116}}
\author[addressref=aff1]{\inits{S. J.}\fnm{Samuel~J.}~\snm{Van~Kooten}\orcid{0000-0002-4472-8517}}
\author[addressref=aff1]{\inits{M.J.}\fnm{Matthew~J.}~\snm{West}\orcid{0000-0002-0631-2393}}

\address[id=aff1]{Southwest Research Institute, Boulder, Colorado, USA}
\address[id=aff2]{NASA Goddard Space Flight Center, Greenbelt, Maryland, USA}
\address[id=aff3]{ADNET Systems, Inc., Greenbelt, Maryland, USA}
\address[id=aff4]{Independent Researcher}

\runningauthor{D.B. Seaton et al.}
\runningtitle{Observing the Corona at an Annular Eclipse}

\begin{abstract}
We present results of a dual eclipse expedition to observe the solar corona from two sites during the annular solar eclipse of 2023~October~14, using a novel coronagraph designed to be accessible for amateurs and students to build and deploy.  The coronagraph (\textit{CATEcor}) builds on the standardized eclipse observing equipment developed for the Citizen CATE 2024 experiment.  The observing sites were selected for likelihood of clear observations, for historic relevance (near the Climax site in the Colorado Rocky Mountains), and for centrality to the annular eclipse path (atop Sandia Peak above Albuquerque, New Mexico). The novel portion of CATEcor is an external occulter assembly that slips over the front of a conventional dioptric telescope, forming a \textit{shaded-truss} externally occulted coronagraph. CATEcor is specifically designed to be easily constructed in a garage or ``makerspace'' environment. We successfully observed some bright features in the solar corona to an altitude of approximately 2.25\,R$_\odot$ during the annular phases of the eclipse. Future improvements to the design, in progress now, will reduce both stray light and image artifacts; our objective is to develop a design that can be operated successfully by amateur astronomers at sufficient altitude even without the darkened skies of a partial or annular eclipse.
\end{abstract}
\end{frontmatter}

\section{Introduction}
     \label{sec:intro} 

The first coronagraph in the United States was at the original site of the High Altitude Observatory, at the top of Fremont Pass, in Climax, Colorado. Conceived by Donald Menzel and initially operated by his 25-year-old graduate student, Walter Orr Roberts, the observatory opened in 1940, and made its first successful observations of the solar corona in October 1941. \citep[See][for a rich and detailed summary of the coronagraph's conception, construction, and initial operations.]{Bogdan2002}

Climax's first coronagraph was an instrument based on the design of the \citet{lyot_1930} coronagraph at Pic-du-Midi, and observed narrowband emission in the coronal red (Fe~\textsc{x} at 637.4\,nm) and green (Fe~\textsc{xiv} at 530.3\,nm) lines. Although these narrowband observations proved extremely challenging, the Climax coronagraph provided some of the first detailed evidence linking coronal activity to geomagnetic disturbances \citep{Shapley1946}.

A decade later, \citet{Wlerick1957} reported on a new instrument, the \textit{K-coronameter}, which made broadband polarimetric observations; these proved much less challenging to make, particularly at the high altitude of the Climax observatory station. During its period of operation, which lasted until July 1972, when the observatory was shuttered in favor of more modern facilities and more favorable locations, Climax's observations led to breakthrough progress in coronal physics and key technological development as well. Of particular interest to us were reports of Climax observations of the corona made in support of research during the 1950 solar eclipse in Alaska \citep{Roberts1951}. The instruments and techniques developed in Climax spawned numerous successor instruments, including space-based coronagraphs that flew on \textit{Skylab} and the \textit{Solar Maximum Mission} \citep{MacQueen1974, MacQueen1980} and as the several coronagraphs hosted at the Mauna Loa Solar Observatory, the most recent of which is the COSMO K-coronagraph \citep{dewijn_2012}.

Although a few coronagraphic instruments have been tested in Boulder in the years since Climax's closure, few if any observations of the corona have been made in Colorado. Although developing dedicated coronagraphs for long-term scientific studies from the Rocky Mountains would present a major -- and potentially infeasible -- undertaking, the solar eclipse of 2023~October~14 presented an interesting opportunity to use high-altitude vantages, like Climax, to attempt to develop a much simpler instrument that could image the corona specifically during the eclipse, when solar brightness is reduced and the sky is significantly darkened.

Our goals for this experiment were threefold: first, because we were developing a distributed network of observing stations for the 2024 total eclipse, \textit{Citizen CATE 2024} \citep[CATE24;][]{Caspi2023}, attempting coronagraphic observations during the annular eclipse would provide an opportunity to test the same equipment and procedures for the total eclipse under realistic conditions. Second, we wanted to develop methods to make simpler coronagraphs that could potentially be built and operated by amateurs in the future. Finally, the endeavor offered a chance to explore the history of coronagraphy in Colorado and share some of the experiences of our scientific predecessors -- to walk in the footsteps of giants.

Though scientifically useful coronal measurements have been made at radio wavelengths during partial and annular eclipses \citep[e.g.,][]{Gary1987, Kathiravan2011} and partial eclipse observations in extreme ultraviolet from space have provided useful opportunities to characterize instrument performance \citep[e.g.,][]{Seaton2013, Goryaev2014}, we are unaware of any successful amateur observation of the corona in visible light from the ground outside of a total eclipse. However, the significant obscuration of the Sun during the eclipse in accessible locales higher than 10,000~ft above sea level, the capability of modern 3D printing to generate the highly tailored optical components required to build an effective coronagraph, and modern detectors capable of straightforwardly measuring polarization provided a unique opportunity to make a novel observation with amateur- or student-accessible tools. 

This paper describes our instrument, which we named \textit{CATEcor} (Section~\ref{sec:catecor}), our field test (Section~\ref{subsec:inital_test}) and eclipse expeditions (Sections~\ref{subsec:co} \& \ref{subsec:nm}), our eclipse observations (Section~\ref{sec:observations}), and the results of our data analysis (Section~\ref{sec:analysis}). We conclude with a discussion of lessons learned and future prospects for additional expeditions and community participatory science (``citizen science'') efforts (Section~\ref{sec:discussion}).

\section{CATEcor Design,  Equipment, and Procedures}
\label{sec:catecor}
CATE24, a next-generation evolution of the originally successful Citizen CATE experiment for the 2017 total eclipse \citep{Penn2017, Penn2020}, will deploy 35 identical sets of equipment to locations all along the path of the 2024~April~8 total solar eclipse to make broadband polarimetric observations of the corona between about 1 and 3\,\Rs. Barring bad weather along the eclipse path, the experiment is expected to return over an hour of continuous coronal observations, which can be used to characterize structure and dynamics in a region of the corona that has historically been under-observed. Because the CATE24 setup was already optimized for the kinds of coronagraphic observations we hoped to make during the annular eclipse, and because a primary goal of our experiment was to test our equipment under realistic conditions, we based the design of CATEcor on the existing CATE24 equipment.

\citet{DeForest2024} describe the CATEcor coronagraph conception, design, and implementation in detail. \citet{Patel2023} discuss the CATE24 setup and demonstrate its capabilities using observations from the 2023~April~20 hybrid-total solar eclipse in Australia. The CATE24 imager consists of:
\begin{itemize}
    \item a FLIR Blackfly BFS-PGE-123S6P-C polarization-sensitive camera,
    \item a Long Perng S500G-A doublet ED refractor telescope with 80\,mm aperture and focal ratio of \textit{f}/6.25, with a DSUV2 UV/IR cut filter, both procured from Daystar Filters, and
    \item an iOptron GEM28 German Equatorial tracking mount and tripod.
\end{itemize}

The FLIR camera uses a Sony IMX253MZR sensor, which provides images with 4096$\times$3000 pixels, where each 2$\times$2 macropixel uses on-chip filters to observe polarization angles of 0\degr, 45\degr, 90\degr, and 135\degr. The optics, coupled with the camera, provide a field of view (FOV) of 1.63\degr$\times$1.19\degr \ (or $\pm$3.05$\times\pm$2.23\,\Rs\ at the time of the eclipse). Preliminary analysis of lab tests on the cameras to be used for the CATE24 total eclipse equipment showed the polarization efficiency of the cameras to be $>90\%$; a detailed analysis is underway and will be presented in a paper covering the CATE24 total eclipse observations.

The 3D-printed CATEcor coronagraph assembly uses a novel ``shaded truss'' design, in which the supports for the external occulter are fully shaded by the occulter and receive no direct illumination from the Sun. The coronagraph assembly is fitted over the front of the CATE24 telescope, and aligned manually using three pairs of adjustable thumb screws. The occulter is a 3D-printed near-cylinder, with a slightly curved surface to minimize diffraction at the occulter edge and permit some tolerance in alignment angle without degrading optical performance significantly. It is supported 75\,cm from the 30\,mm assembly aperture by a shaded hexapod structure composed of six 2-mm carbon fiber rods. An adjustable iris permits the aperture to be stopped down to achieve different degrees of occultation, though we ultimately observed during the eclipse with the iris fully open.

\begin{figure}    
\centerline{\includegraphics[height=0.3\textheight,clip=]{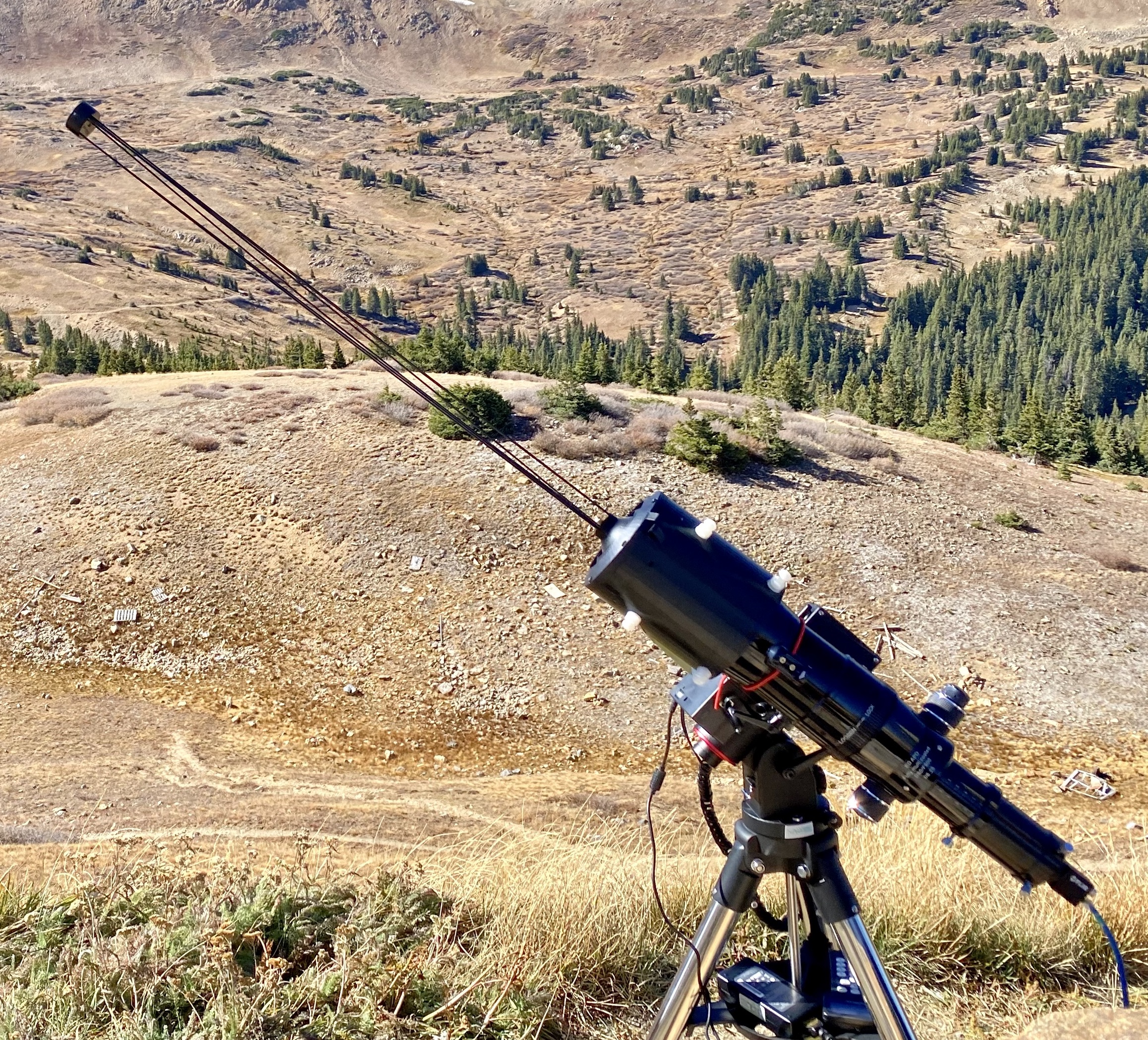}  \includegraphics[height=0.3\textheight,clip=]{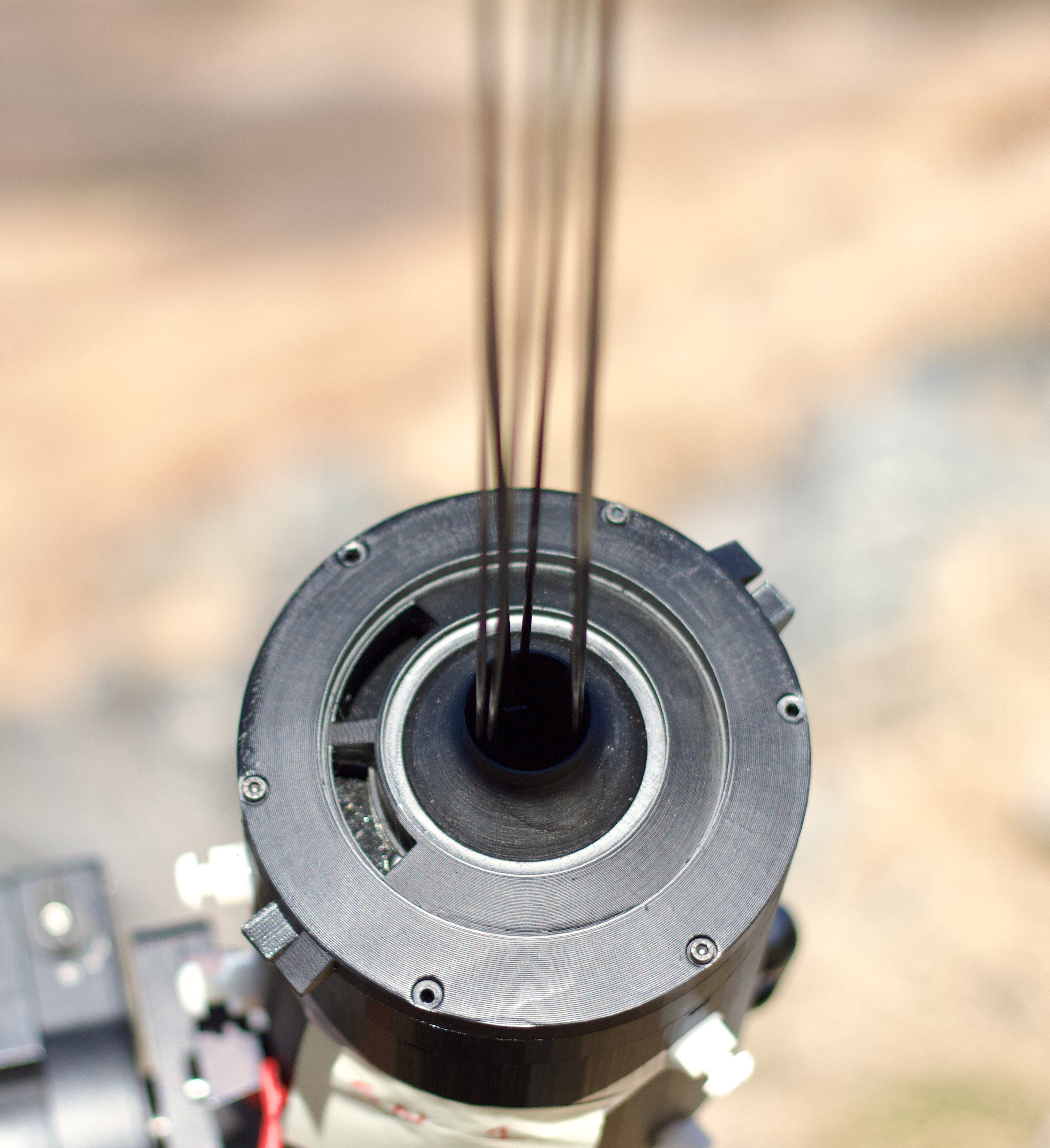}}
\small
\caption{Overview of CATEcor coronagraph assembly and CATE telescope setup during field tests on Loveland Pass in Colorado (left). CATEcor assembly aperture in the coronagraph occulter umbra during the same field test (right).}
\label{fig:CATEcor_overview}
\end{figure}

Figure~\ref{fig:CATEcor_overview} shows an overview of the CATEcor assembly mounted on the CATE24 telescope from our first field test, as well as a close-up of the shaded aperture opening and associated baffles. 

The telescope and mount must be manually aligned during observatory setup, and although the mount supports alignment to the pole star, because of the limitations of when we could observe at some sites, alignment was also achieved by dead reckoning using spirit level, inclinometer, and compass readings. We were able to achieve good, but not perfect, tracking of the Sun. Occasional adjustments were required to maintain pointing. We adjusted the occulter position by eye by using the adapter thumb screws and observing the position of the occulter shadow relative to the Sun in images from the telescope assembly. Observations were managed by custom software developed specifically for the CATE24 project.

CATE24 is intended to achieve high-dynamic-range (HDR) imaging of the corona by acquiring and combining a set of eight logarithmically-spaced exposures (0.123--399.995\,ms) to cover the three orders of magnitude of dynamic range of the corona in our FOV. We used these default observing sequences. However, as we discuss in Section~\ref{sec:analysis}, we analyzed only the 1.3-ms exposures.

We obtained dark and flat images for calibration both before and following each campaign. Darks use the same exposure times as the primary observations and are made at ambient temperature, and are achieved by pointing away from the Sun and covering the telescope aperture. We obtained flats using two methods intended to capture not only anisotropies in the detector, but also the coronagraph vignetting function. First we took sky flats by obtaining images while pointing the assembly several degrees away from the Sun. To characterize the importance of any artifacts due to the residual polarization of the sky, we pointed in multiple directions away from the Sun during this process. We also took flats by placing a diffuser plate directly in front of the occulter. However, in the absence of a collimated light source behind the occulter (i.e., the Sun) the most important artifacts observed in the flats varied significantly from our eclipse images, and thus were of limited value in correcting our images.

We tested focus before and after attaching the occulter using images of the solar limb and sunspots, obtained by observing through a neutral density solar filter while the coronagraph assembly was not attached to the telescope. Sunspot images served a dual purpose, allowing us to determine the orientations of our images with respect to solar north via comparison with data from other observatories. (The cameras were aligned to celestial north to make pointing adjustments on our German Equatorial mount more intuitive, but the orientation was set by hand, so images were necessary to address any small error in camera alignment.)

\section{Field Expeditions}
\label{sec:expeditions}

We deployed CATEcor into the field on three separate occasions: first, a field test the week before the eclipse to test observing procedures and characterize CATEcor's performance under realistic conditions. The test allowed us to identify problems with the instrument performance itself as well as key challenges for its operation to help refine procedures to ensure smooth operation on eclipse day.

We deployed to two locations on eclipse day: one at Loveland Pass, Colorado, where there was only a partial eclipse but we could observe from very high altitude, and one at Sandia Crest, New Mexico, where the eclipse was fully annular, but the altitude somewhat lower. We discuss each of these campaigns separately below.

\subsection{Initial Test: Loveland Pass, Colorado}
\label{subsec:inital_test}

We planned an initial test of the CATEcor instrument about a week before the eclipse, on October~6, at Brainard Lake Recreation Area, at an altitude of about 10,300~feet above sea level, to the west of Boulder, Colorado. However, cloudy conditions along the Front Range required a change in plans. Loveland Pass was the next closest easily accessible high-altitude location, at 12,000 feet, and so we selected it for our field test.

The primary goal of this initial campaign was to demonstrate all required equipment, validate our procedures for the eclipse observation, and obtain an inital assessment of the performance of the CATEcor coronagraph assembly and CATE24 instrumentation. During this test we had to contend with the full Sun, rather than a small annulus or narrow crescent, and maximum sky brightness, so the instrument performance would be an upper limit -- in contrast to the darker skies and much-reduced sunlight during the eclipse.

\begin{figure}
\centerline{\includegraphics[width=1.0\textwidth,clip=]{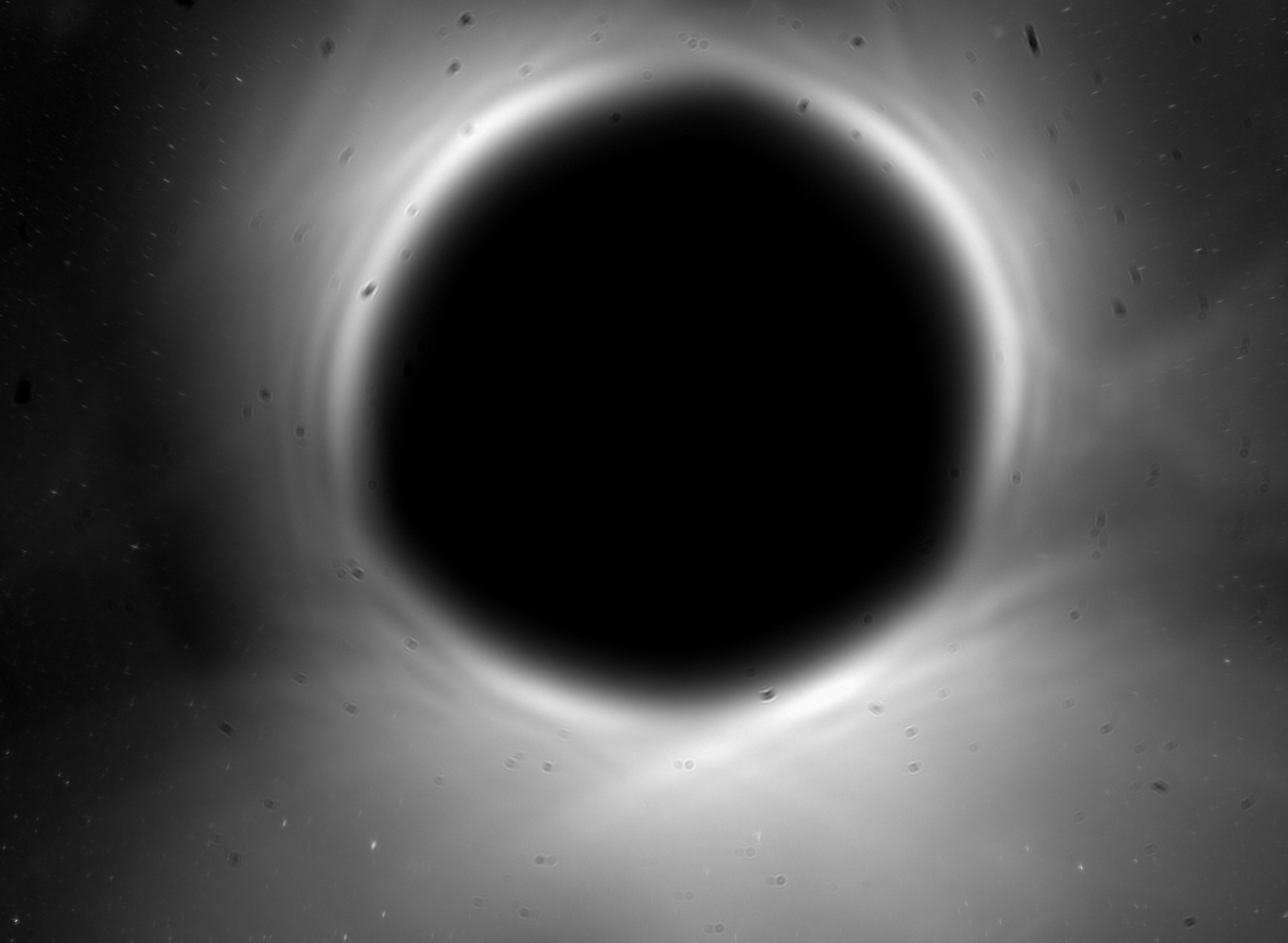}}
\small
\caption{Square-root-scaled 1.3-ms total brightness image of the CATEcor occulter from our first field test. In addition to the occulter itself, there are numerous artifacts, apparently due to light scattering off of different surfaces on the coronagraph assembly. Bright points in the image are illuminated dust blowing above ground level.}
\label{fig:field_test}
\end{figure}

Figure~\ref{fig:field_test} shows a sample of the data obtained during the field test, square-root scaled to address the relatively large dynamic range of the image. The Sun is hidden behind the occulter in this image, and we interpret most of the brightness of the image to be diffraction at the occulter or scatter off surfaces in the coronagraph assembly. The air was full of dust particles, brightly illuminated by the Sun, during the test, which appear as small, bright spots throughout the image. 

We tested two occulter assemblies during this trip. One turned out not to be sufficiently stiff to maintain the alignment of the telescope and coronagraph, particularly because the day was windy. (Although we had attempted to set up our equipment in a sheltered location, the wind shifted as the afternoon went on.) The second assembly was stiff enough for the occulter to remain relatively stationary, even in the wind. Although the occulter shape was optimized to suppress diffraction, the unpainted black polyethylene terephthalate glycol (PETG) material of the occulter and assembly, as well as the carbon fiber truss supports, were reflective enough to scatter light in complex and unexpected ways, such as we see in the diagonal, cross-hatched pattern near the bottom of the dark occulter image.

The highly structured scattered light we observed in these initial images motivated a key change to the design: we applied a thin coat of Krylon Ultra-Flat Camouflage Black spray paint to the assembly to suppress glint. (A comparison of painted and unpainted occulter performance images from pre-eclipse testing appears in Section~\ref{sec:observations}; a detailed discussion of the paint application appears in \citealt{DeForest2024}.)

Two important lessons learned during the field test both related to the initial alignment of the occulter assembly and our ability to maintain this alignment in the field. Aligning the occulter by hand proved very difficult, but alignment by slightly adjusting the telescope pointing was more straightforward. Therefore, we only coarsely aligned the occulter assembly to the telescope by hand, and then adjusted the instrument pointing to isolate the Sun behind the occulter. This meant sacrificing the precise centering of the Sun and occulter in the FOV, but significantly reduced the time to achieve alignment.

Maintaining alignment requires both accurate tracking of the Sun and improving the occulter performance in the wind (or, at a minimum, isolating the instrument from the wind). \citet{DeForest2024} describe how we improved the overall stiffness of our second generation of occulters to improve their performance in windy conditions. We also sought out better-sheltered observing locations for the eclipse, and refined our telescope alignment procedure to achieve the best possible tracking of the Sun, including using careful compass and GPS readings to set the telescope's axis alignment, and careful leveling to ensure the right latitudinal settings for the mount.

We also refined data collection and calibration sequences, which had to be adapted from those used during a total eclipse to account for the differences resulting from the use of an occulter. This included observing sunpots to determine the camera's orientation with respect to solar north and collecting flats using several different methods both before and after the eclipse. We also observed that dark current was not a significant concern, because under the cold observing conditions in the field the camera produced $<$1~DN/s/pixel of average dark current, and all of our exposures were much less than 1~s, though dark+bias images are still required to remove the image offset resulting from camera bias.

\subsection{Partial Eclipse: Loveland Pass, Colorado}
\label{subsec:co}
We selected Pass Lake, at an altitude of $\sim$11,800 feet, just south of, and about 200 feet below, the summit of Loveland Pass, as the first of our observing sites. Originally we had targeted the old site of the Climax Observatory, on Fremont Pass near Leadville, Colorado, about 600~feet lower, largely for historical reasons. Unfortunately, the Climax site is owned and managed by the Climax Molybdenum Mine, which declined to allow our team access to the area. 

Pass Lake provided an easily accessible high-altitude observing site with ample flat ground to set up our equipment, some shelter from wind, and good sight lines for the eclipse. It also seemed appropriate, as Janet Roberts, Walter Orr Roberts' wife, had written evocatively of her first impression of the Rocky Mountains, driving over Loveland Pass -- at the time one of the only routes over the Continental Divide in the Front Range -- on the Roberts' first trip to Climax in 1940.

\begin{quote}
    I had never in my life seen such mountains, let alone driven in them. The road [over Loveland Pass] was a narrow, graveled two lanes, which climbed relentlessly, doubling back on itself in sharply angled switchbacks.... My nervousness was heightened by the steep drop-off of the mountain on one side and the gouged-out cliffs that shadowed the road on the other....
    
    But the weather was fine, and just over the summit (the marker read ``11,992 feet above sea level'') we found a small snowfield left from the winter before. All three of us paused to throw snowballs in July and to look in wonder at the broad stretches of arctic tundra carpeted with wild-flowers and the grey, craggy peaks above our heads. \citep{JRoberts1989}
\end{quote}

\begin{figure}
\centerline{\includegraphics[width=1.0\textwidth,clip=]{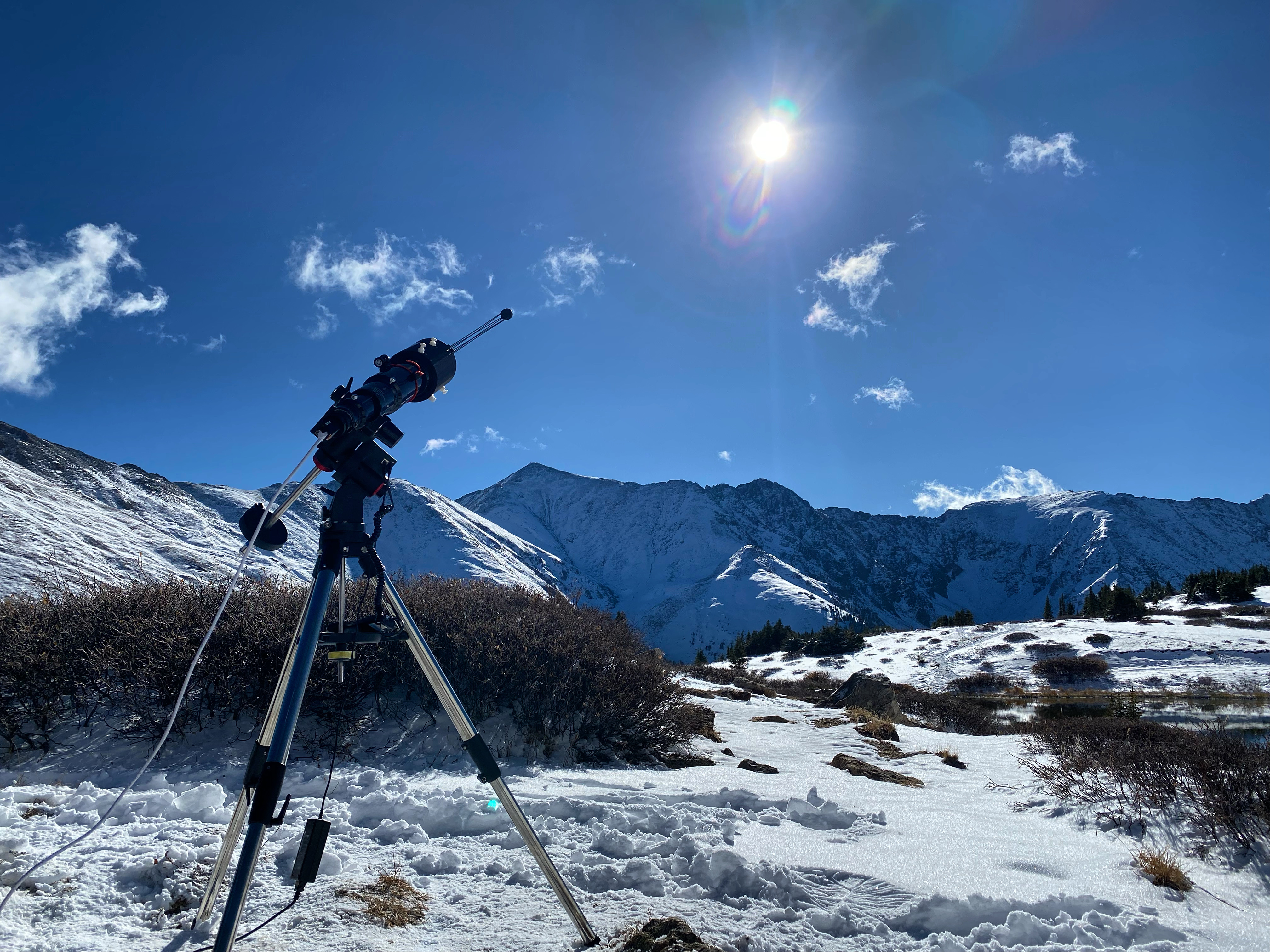}}
\small
\caption{The CATEcor telescope and coronagraph assembly at Pass Lake, just below Loveland Pass, in Colorado, during the early phases of the 14~October~2023 partial eclipse. Our observing site offered completely unobstructed views to the south, over the Arapaho Basin Ski Area. A snowstorm on 12~October complicated working conditions at the site, but helped to suppress airborne dust.}
\label{fig:co_location}
\end{figure}

A large snowstorm in the Colorado mountains on 12~October left several inches of snow on the ground at our site, which made footing a challenge, but significantly reduced dust in the air from what we observed during our dry run a week earlier (see Figure~\ref{fig:co_location}). Low temperatures both on eclipse day and during our pre-eclipse calibration campaign on 13~October were well below $-5${\degr}C. (Note that this is somewhat below the Blackfly camera's reported range of operating temperature, 0--50{\degr}C, but we nonetheless encountered no problems operating the camera in these cold temperatures.) Skies both on 13 and 14~October were largely clear with excellent seeing, with a strong breeze out of the southeast on eclipse day.

From Loveland Pass, the eclipse reached a maximum of 80.37\% at 16:35:07~UT (10:35~AM MDT) at an altitude of 32.9\degr\ in the southeast sky. First contact occurred at 15:13:14~UT and last contact at 18:04:51~UT; sunrise over the mountains to the south was at about 08:00~AM local time (14:00~UT). 
   
Our dry run the day before the eclipse had allowed us to refine our setup, solar acquisition, and observing procedures, so we were able to work efficiently in very cold temperatures (below $-10${\degr}C early in the morning) to ensure we had acquired the Sun and could observe regularly beginning at first contact. Because annularity was not visible from our Colorado site, there was no strongly preferred time for observations. We therefore took data throughout the entire eclipse, increasing the frequency of observations near the peak of partial coverage.

Our dry run had also demonstrated that accurate telescope alignment could be achieved without stellar observations, using careful compass sightings to align the telescope mount to true north, and setting the altitude by hand to match the 39.655\degr\ latitude of our observing site. Because of the cold, snowy, slippery conditions at the observing site, setting up on eclipse day in daylight significantly reduced the complexity of the observation campaign, and reduced the risk of a mishap occurring during setup in dark conditions.

On eclipse day, we ran the nominal CATE24 observing sequence with its eight exposures, but found that, due to the bright ring of diffraction and scattered light from the occulter, a single 1.3\;ms exposure was sufficient to capture the full field of view with neither saturation nor underexposure. However, because the observing sequence could not be customized easily on the fly, we proceeded with the full sequence throughout the eclipse. We took bursts of roughly 100 images every five minutes through the early eclipse, and every 2.5 minutes in a roughly 30-minute window around the time of totality.

Although remote, our observing site turned out to be a nonetheless popular eclipse viewing area, and we handed out many eclipse viewing glasses and pinhole projectors to visitors, in addition to explaining our experiment and conducting a live-streamed internet broadcast from our site explaining the CATEcor project, its goals, and providing an introduction to the CATE24 total eclipse project. (We note that, compared to the number of visitors seeking a good view of the eclipse, there were relatively few backcountry skiers seeking to enjoy the early-season snowfall. The handful who stopped by were surprised to learn there was an eclipse in progress -- though a few had noticed the unusual light quality on a snowy fall morning.)

\subsection{Annular Eclipse: Sandia Crest, New Mexico}
\label{subsec:nm}
We selected Sandia Crest, just east of Albuquerque, New Mexico, as the second of our observing sites. Sandia Crest was a near-ideal observing site, with an altitude of 10,600~feet and placement near the centerline of the annularity. The specific spot was La Luz Grand Enchantment Trailhead, just south of the crest parking lot at the top of New Mexico State Highway~536.  

Sandia Peak is characterized by steep cliffs to the southwest, interrupting a gentle rocky slope on the northeast side of the crest ridge. The cliff provides a sharp boundary layer between the prevailing southwesterly winds and a nearly-still lee bubble on the ridge itself.  The eclipse, on 14~October, occurred during the annual Balloon Fiesta in Albuquerque, and close to dawn we were able to observe the famous ``Albuquerque box'' of 
counterflowing katabatic northerlies in the valley and thermally driven local southerlies aloft, by tracking the mass of hot-air balloons launching from the fiesta field nearly a vertical mile below us.

Notably, even at a well-traveled observing site near a mid-sized city, access was a concern: during final planning, we noticed that New Mexico State Highway~536, the sole motor access route to Sandia Crest, was scheduled for maintenance and partial closure on the day of the eclipse, and flagged as intermittent-access only. A phone call to the New Mexico Department of Transportation, two weeks prior, ensured that the work was suspended on that day, allowing full access for us and for anyone else in the area.

As in Colorado, we dry-ran the observations on 13~October, so as to be fully rehearsed and ready on eclipse day. Unlike in Colorado, we arrived well before dawn on both days to align the telescope to the North Star. We were treated to the joint display of both Venus in the East and Jupiter just west of zenith in the frigid early 
morning dark. Ambient temperature was near 0{\degr}C pre-dawn, but rapidly rose by 25{\degr}C between dawn and eclipse first contact, shortly after 10~AM local time. This shift in temperature drove our observing requirement to focus the telescope on sunspots through a solar filter just before the eclipse, then to replace the solar filter with the CATEcor assembly for the actual observation.

\begin{figure}
\centerline{\includegraphics[width=1.0\textwidth,clip=]{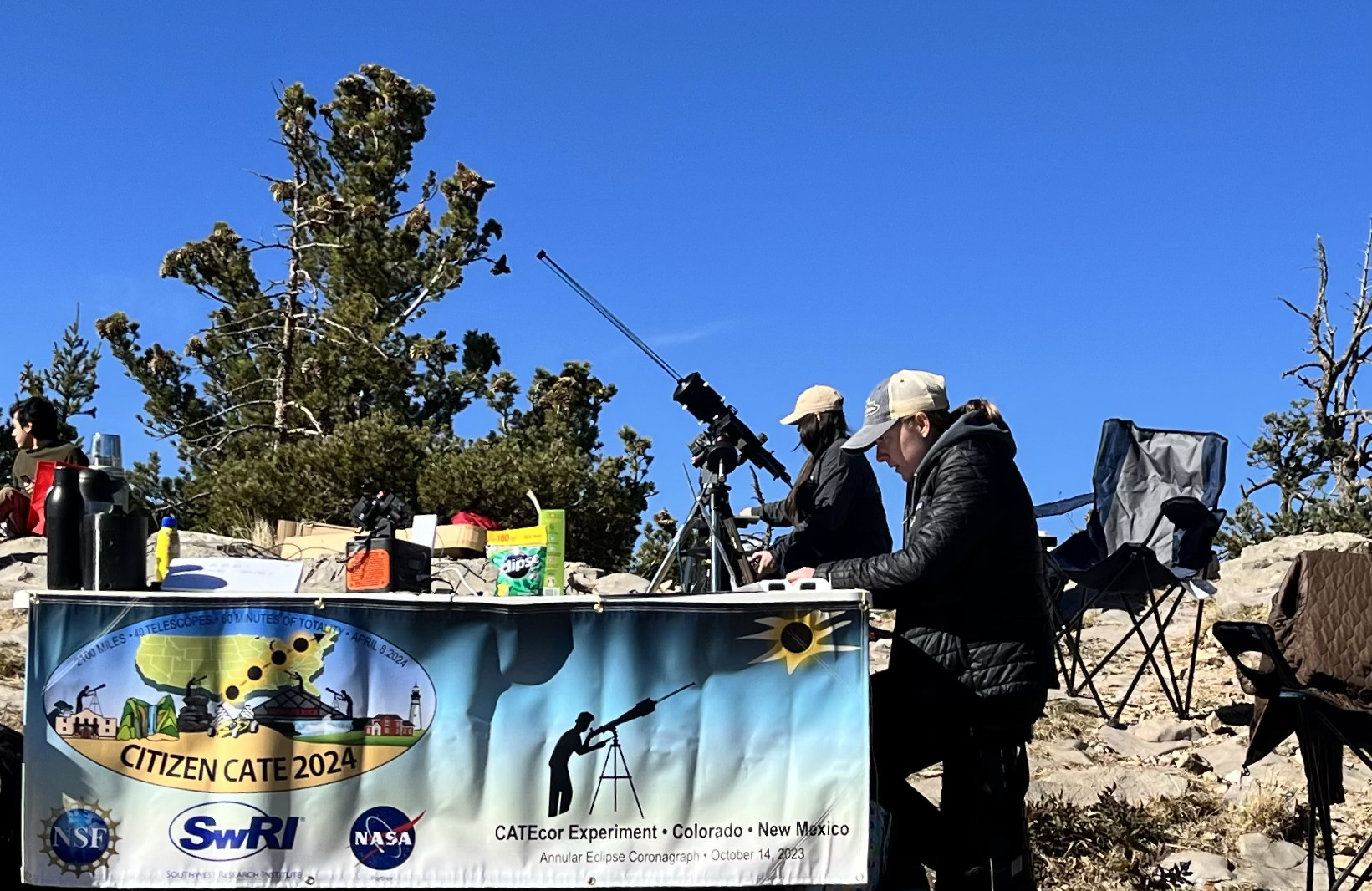}}
\small
\caption{New Mexico CATEcor station tracks the Sun at La Luz Grand Enchantment Trailhead at Sandia Crest, NM, on 2023-Oct-14 in the late partial phase of the day's annular eclipse.  The ``horizon'' behind the telescope is the lip of a cliff and steep slope extending to the valley floor more than 1500 meters below. Co-authors SK and SD are operating the telescope.}
\label{fig:nm_location}
\end{figure}

The ridge ground at La Luz Grand Enchantment Trailhead is pocked and layered limestone, tilted by upthrust. While it made for awkward footing, the ground was perfect for setting up a telescope tripod as the natural depressions in the stone anchored the tripod feet (see Figure~\ref{fig:nm_location}). At 1.5~m above the ground, and 3~m back from the sharp cliff edge, the air was nearly still, although the prevailing wind outside that lee bubble was 8--15~knots with occasional stronger gusts.

As deployed, the telescope was roughly 1~m below the 0.5--1~m boundary layer between the prevailing 
wind and the lee shelter of the Sandia Crest face. Turbulent eddies did affect CATEcor, shaking the 
occulter visibly in the collected data; but the proximity of the boundary layer and the short focal length of the telescope meant that seeing effects were essentially invisible in the actual data.

Sandia Crest was a very popular destination that day, and we had planned for significant contact with the public. We set up a folding table and a banner describing the experiment, and had on-hand roughly 500 pairs of disposable eye-safe ``eclipse shades'' which we handed out to passersby, together with PUNCH pinhole projectors \citep{Morrow_etal_2023} and Eclipse brand chewing gum.  We also set up publicly-accessible solar observing equipment including an H-$\alpha$ telescope and an eyepiece-projection rig using a pair of binoculars, and brought extra personnel who could talk with the public to avoid disturbing the actual observation. This role rotated between CD, NE, SO, and AD as needed. The tilted, rough ground made a display pavilion impractical; fortunately, the clear weather and the cliff lee conditions also made it unnecessary.

At our Sandia Crest site, the partial eclipse began at 15:13:18~UT (09:13~AM Local Time), 
second and third contact occurred at 16:34:39 and 16:39:29~UT, respectively, with the peak of annularity at 16:37:03~UT (10:37~AM local time), reaching a maximum coverage of 89.56\%. The end of the eclipse (fourth contact) occurred at 18:09:33~UT. The Sandia Crest team concentrated their efforts on observing during annularity specifically, when suppression of the background solar brightness was at a maximum.

\section{Observations}
    \label{sec:observations}      

\begin{figure}
    \centering
    \includegraphics[width=1.0\textwidth,clip=]{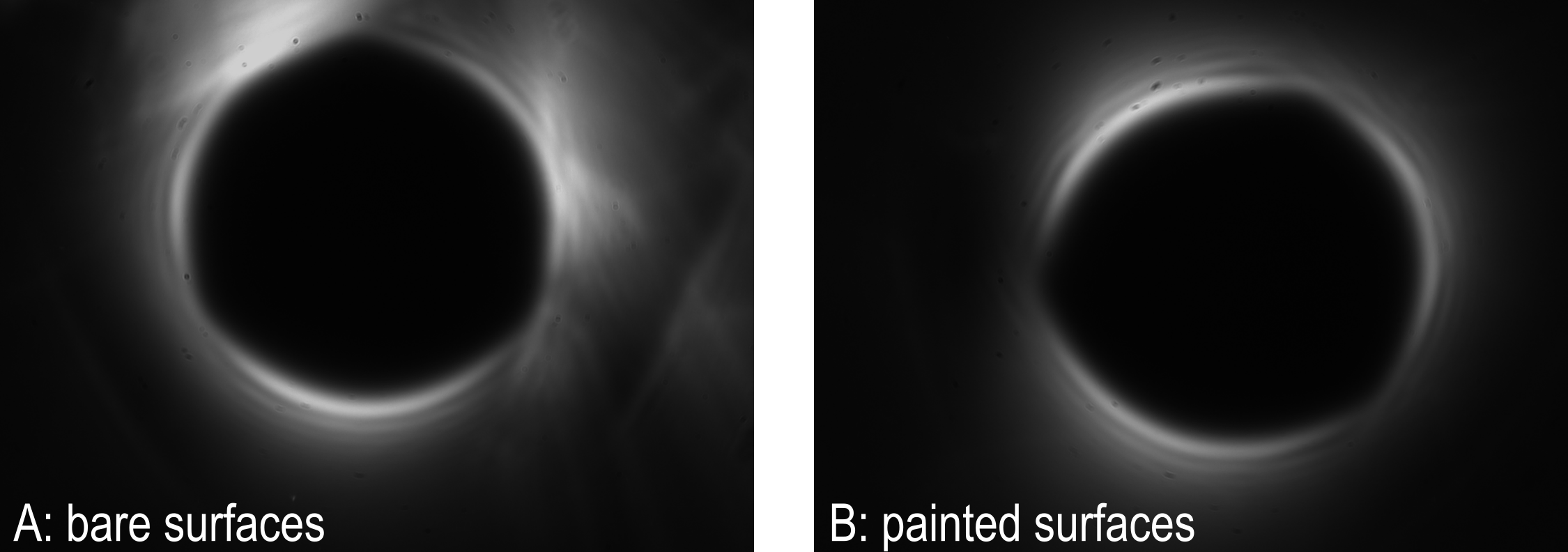}
    \caption{Side-by-side comparison of stray light patterns captured by (A) unpainted and (B) painted occulter assemblies in full Sun at the Sandia Crest location shows the benefits of treatment with flat black paint to prevent glint.}
    \label{fig:black-paint}
\end{figure}

As we discussed above, at both sites we performed a ``dry run'' on 13~October, the day before the eclipse. Teams arrived early and aligned the German equatorial telescope mounts based on compass readings for true north, taking latitude and local magnetic variation into account. Each crew brought both an unpainted, raw 3D-printed occulter assembly and an occulter assembly painted with Krylon Camouflage black non-reflective paint to test the relative performance of the two configurations. Although we had some concerns that the black paint would detrimentally alter the surface figure of the occulter, our head-to-head test showed that the paint significantly reduced artifacts resulting from reflected light on the relatively shiny plastic of the unpainted occulter, and we selected the painted occulters to use on eclipse day (Figure \ref{fig:black-paint}).  

We collected images using the same observing sequence (``totality scan'') as planned for the CATE24 total-eclipse experiment, and as described in Section~\ref{sec:catecor}.  

CATE24's camera software returns 12-bit-per-pixel images in 16-bit TIFF format, with 4 bits of padding in the least significant bits, with polarization information encoded in $2\times2$ groups of pixels: images must be bit-shifted to obtain detector counts and de-interlaced to obtain each separate polarization image. For such short exposures, dark current was negligible, so during calibration we only removed the camera bias from each image, neglecting dark current effects.

We tried several methods to obtain flat field images, including taking images of the sky and a glass diffuser with the coronagraph assembly attached to the telescope, and a set with the diffuser attached directly to the telescope without the coronagraph assembly. We pointed to several positions on the sky to attempt to obtain images where we could separate the polarization of the sky from the overall performance of the telescope and camera.

Our hope was that flat field images obtained with the coronagraph assembly attached would help correct for the instrument vignetting function, but we found that neither of our two sets of images provided a consistent measurement of both camera flatness and instrumental vignetting function, so we elected to neglect the latter effect during image calibration. The diffuser-only flats provided an adequate characterization of anisotropy from pixel-to-pixel in the camera, and we used these to flatten the camera response for each image. A few dust particles in the telescope led to artifacts resulting from local imaging of the instrument aperture throughout the images, but the difference between the aperture with the Sun behind the occulter and without the occulter assembly was too significant for the diffuser flats to correct, and these artifacts remain in the images.

\begin{figure}
\centerline{\includegraphics[width=1.0\textwidth,clip=]{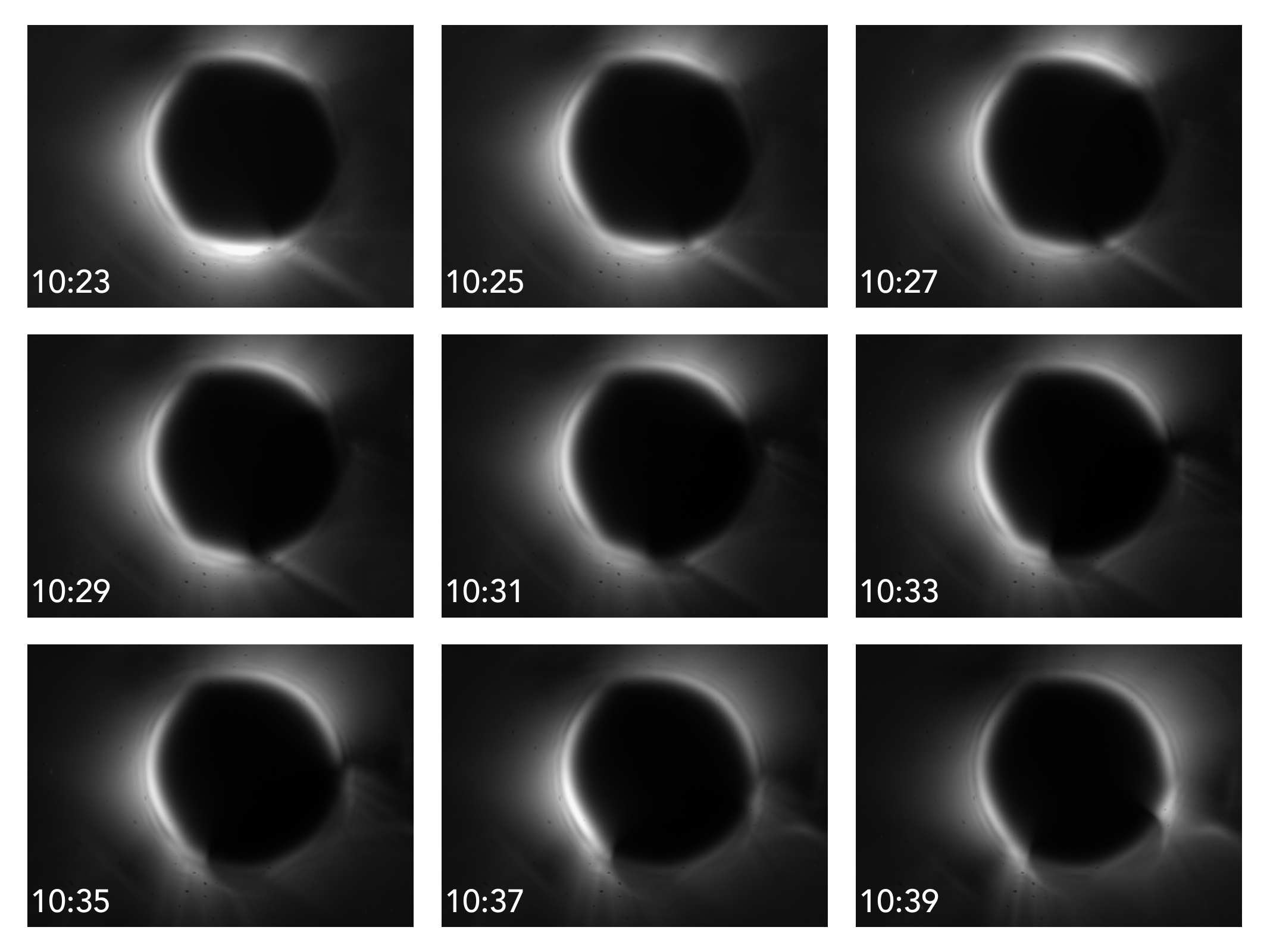}}
\small
\caption{Occulter images showing changes in the background brightness around the time of peak eclipse, in a single polarization channel from our Colorado site. Solar north is approximately 27\degr\ counter-clockwise from vertical.}
\label{fig:colorado_progression}
\end{figure}

We found that the progression of the eclipse led to complex and rapidly evolving artifacts, resulting from the narrow crescent of the Sun illuminating the coronagraph. Figure~\ref{fig:colorado_progression} shows how the pattern of background brightness evolved as the eclipse progressed at Loveland Pass, highlighting how the changing crescent Sun caused changes in the overall background brightness we observed. Because there was no annularity in Colorado, the Sun always appeared as a significantly asymmetrical crescent there, and the images showed a strongly anisotropic pattern of stray light as a result.

Our hope was that the changing background brightness from the occulter might permit the separation of stray light resulting from diffraction at the occulter edge and other reflections from the corona itself, but due to the complexity of the changing pattern this work is still ongoing. For our preliminary analysis, we therefore focus on the observations from Sandia Crest, where, during annularity the stray light in the images was distributed more isotropically around the occulter, resulting in fewer artifacts within the images and simplifying the analysis and interpretation of the data. Additionally, with a factor of two more obscuration of the Sun, the sky brightness was an equivalent factor darker as well.

\section{Data Analysis}
  \label{sec:analysis}
The key goal of the CATEcor experiment was to determine whether it is possible to detect the corona using our relatively simple, purpose-built coronagraph during an annular eclipse, and, ultimately, develop a prototype instrument that amateur observers might one day be able to use to make meaningful measurements of the polarized corona outside of an eclipse, provided they can observe from sufficient altitude. As a result, our data analysis focuses on determining if the corona is visible in our observations, rather than on the detailed appearance or state of the corona at the time of the annular eclipse.

In spite of the eclipse suppressing $\sim$90\% of the sunlight at the peak of annularity, the primary difficulty in detecting the corona is still overcoming the stray light induced by diffraction at the occulter edge and scattering/reflection within the occulter assembly and telescope system. One way many coronagraphs overcome this challenge is to remove ``background'' images by tracking the minimum brightness across the field-of-view over some sufficiently long period that the changes in the corona become separable from static contributions from stray light and the F-corona \citep[one example of this is LASCO's monthly minimum images; see][]{Brueckner1995}. This method is remarkably effective when consistent, long-term observations of the corona are available, but was not feasible for CATEcor, which operated only for a few hours on one specific day. As a result, determining whether the corona is visible above the stray light and sky background requires more sophisticated analysis.

However, three separate pieces of evidence suggest that our experiment did detect the corona. First, CATEcor, like the earliest coronagraphs \citep{lyot_1930}, measures polarized light. To first approximation order, stray light from the sky and the (fainter) F-corona are unpolarized within a few degrees of the Sun, and therefore images created with the observable quantity $pB$ (polarized brightness) rather than $B$ (total brightness) reject most of the sky background even on a non-eclipse day. 

\begin{figure}
\centerline{\includegraphics[width=1.0\textwidth,clip=]{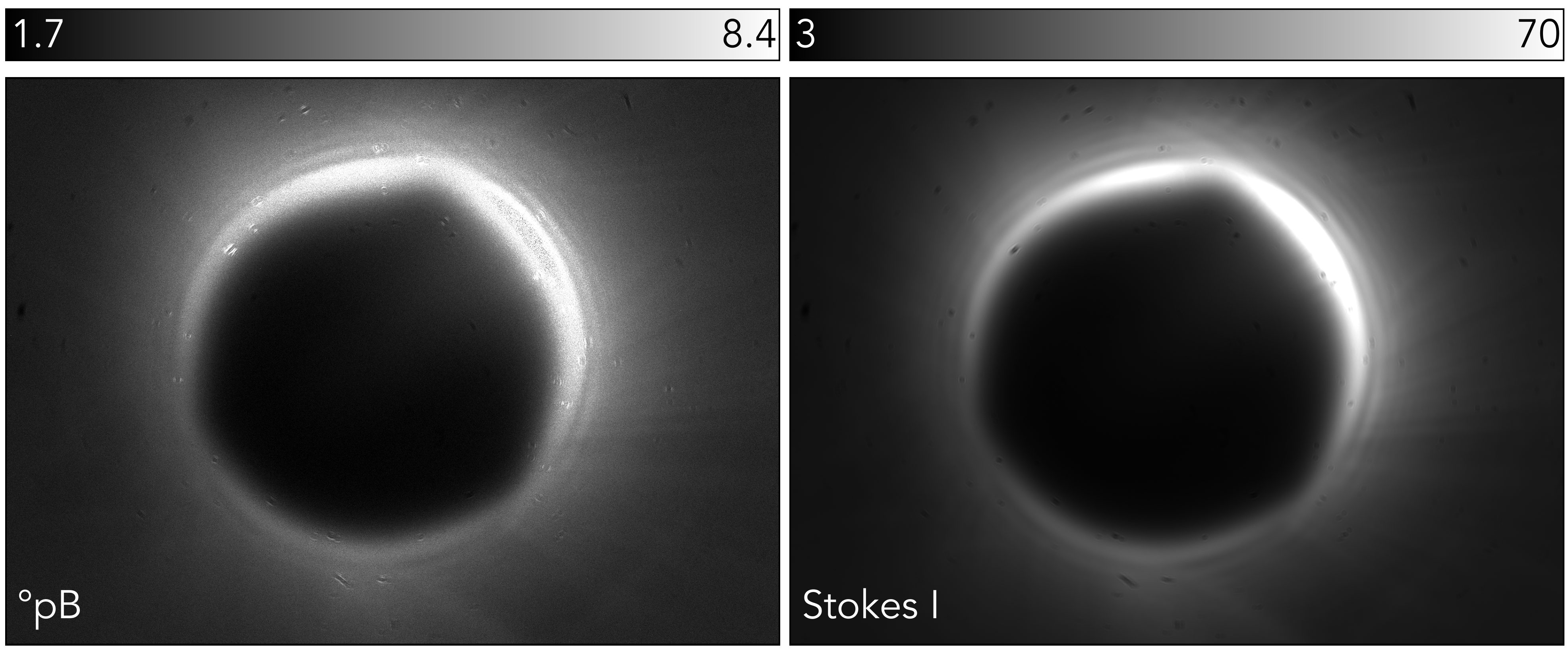}}
\small
\caption{Total polarized brightness ($^{\circ}pB$) and total intensity (Stokes~I) in coronagraph images from the time of peak eclipse in Albuquerque. Comparison reveals the polarization fraction in the image is about 5\% near the edge of the occulter near 1.5\,R$_\odot$, increasing to about 10\% at the edges of the image at approximately 3\,R$_\odot$.}
\label{fig:circ_pb}
\end{figure}

We begin by computing the total polarized brightness, $^{\circ}pB$, in images obtained around the time of peak eclipse in Albuquerque \citep[see the Appendix of][for a complete discussion of the traditional definition of $^{\circ}pB$ and other systems for characterizing polarized intensity in the corona]{DeForest2022}. Figure~\ref{fig:circ_pb} shows a comparison of total intensity and polarized brightness, displayed using square-root scaling; the $^{\circ}pB$ image has a slower overall falloff of intensity and more structure visible outside of the bright diffraction pattern at the occulter edge.

Computing the ratio of polarized brightness to total intensity gives the \textit{polarization fraction}, that is, the fraction of total observed light that is polarized. Here the polarization fraction increases from about 5\% at the occulter's edge, near 1.5\,R$_\odot$, to about 10\% at the edges of the image at approximately 3\,R$_\odot$. Excluding contributions from stray light, we might expect the polarization fraction for observations during the eclipse near 1.5\,R$_\odot$ to be about 50\%, dropping to $<$1\% at 3\,R$_\odot$, considering contributions from the K- and F-coronae and the unpolarized sky background \citep[see][Figure 1]{DeForest2024}. 

However, CATEcor images contain a significant contribution from stray light, which originates directly from diffraction at the occulter edge, and subsequently from scattering of diffracted light by the telescope optics and falls off with increasing distance from the occulter. Therefore, a polarization fraction of a few percent near the occulter edge suggests at least some contribution from the polarized K-corona. Furthermore, the K-corona itself increases its polarization fraction with altitude, up to elongation angles of 2--3\,$R_\odot$ from the Sun, due to the local scattering geometry: the scattering plane of any one photon may be far from the ``average'' scattering plane of the photospheric light as a whole \citep[e.g.,][]{Billings_1966,DeForest_1995}, so our observed increase in polarization fraction with radius is to be expected.  

\begin{figure}
\centerline{\includegraphics[width=1.0\textwidth,clip=]{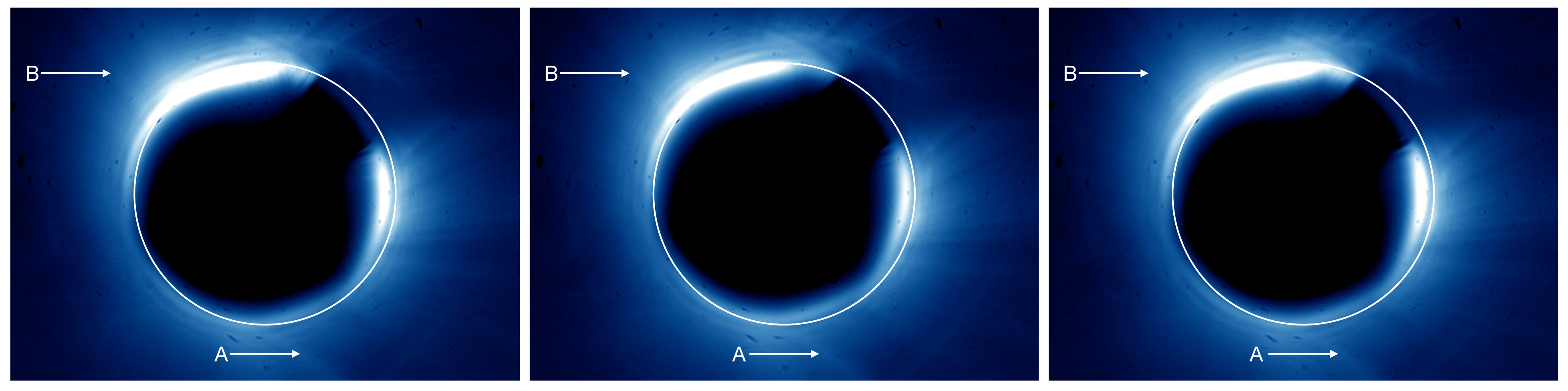}}
\small
\caption{Three sequential snapshots from the Albuquerque eclipse observation. The superimposed circle and arrows are all drawn at the same location with respect to the camera frame in each image and highlight features that move due to occulter wobble (highlighted by the circle and arrow A) and features that remain fixed regardless of occulter motion (arrow B). An animated version of the complete eclipse entrance observation, in which the motion is more evident, is also available.}
\label{fig:occulter_motion}
\end{figure}

A second piece of evidence that we detected the corona resulted serendipitously from the fact that the occulter on the Albuquerque CATEcor setup wobbled significantly during the eclipse. As a result, the stray light pattern resulting from diffraction and instrumental scattering and reflections moved with respect to features fixed on the background sky. Figure~\ref{fig:occulter_motion} and its accompanying animation highlights this effect. In each of the three panels in the figure the overlaid circle and two arrows are in fixed locations relative to the camera frame. Using the circle as a reference point, it is evident that the occulter and associated diffraction pattern change positions in the three successive panels (this motion is even more evident in the animated version of the figure). The same is true for the more or less linear feature indicated by arrow A. Arrow B, however, points to a feature that \textit{does not} move from frame to frame. 

During the observation, the telescope, attached to a heavy tripod and mount, was stable, and thus its pointing did not vary relative to distant objects on the sky. The occulter, supported by thin, flexible carbon fiber rods, moved. Features that move with the occulter therefore must be generated at the occulter or as the result of diffracted light at the occulter edge. This disparity allows us to determine which features in the observations originate inside or outside of the telescope system. The feature labeled with arrow B is one such example, and is likely to be a coronal feature on the sky. As we will see below, these features also appear to correspond well to coronal structures observed by other instruments as well, further suggesting they originate in the corona.

This bring us to the third bit of evidence that CATEcor successfully detected the corona during the annular eclipse: there is general agreement between features observed in processed CATEcor images and contemporaneous observations of the corona.

\begin{figure}
\centerline{\includegraphics[width=1.0\textwidth,clip=]{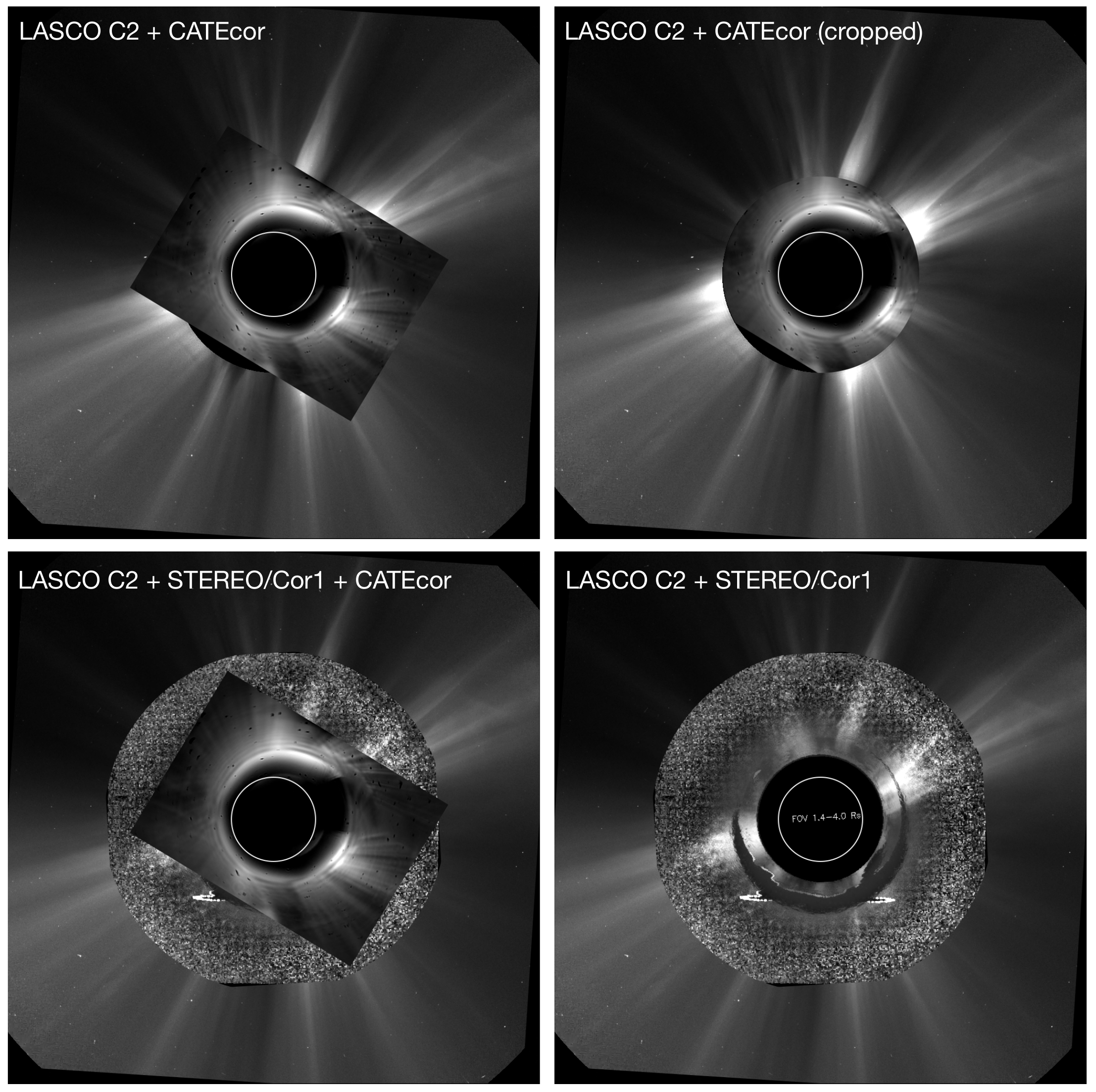}}
\small
\caption{Comparison of radially filtered CATEcor images and observations of the corona from SOHO's LASCO C2 coronagraph (top panels) and STEREO's Cor1 (bottom panels).}
\label{fig:overlays}
\end{figure}

Figure~\ref{fig:overlays} shows a comparison between processed CATEcor images from the Albuquerque annularity sequence and observations from LASCO's C2 coronagraph and STEREO's SECCHI Cor1 coronagraph \citep{Howard2008}. The CATEcor images are processed with an azimuthally varying radial filter \citep{Seaton2013}, which enhances image contrast and improves the visibility of the structure in the CATEcor images (at the expense of photometric meaning). 

Unfortunately, at the time of the eclipse, in October~2023, the availability of visible-light coronal images that fully overlapped with the CATEcor field of view was severely limited. The Mauna Loa Solar Observatory's K-Cor instrument remained offline following the volcanic eruption of November and December~2023 and, to our knowledge, the only other operational coronagraph with a field of view extending inside of about 2.5\;R$_{\odot}$ was the heavily degraded Cor1 coronagraph \citep{Howard2008} on board the STEREO mission. Fortunately, at the time of the eclipse, STEREO-A was separated from Earth by only about 4.8\degr\ heliographic longitude, so it shared a very similar view with CATEcor. To improve the visibility of coronal features in the degraded Cor1 images, we averaged data over consecutive 6-hour blocks from October 13 to the end of October 14 and applied the normalizing-radial-graded filter \citep[NRGF;][]{Morgan2006} filter to improve the contrast of large-scale features. To reduce high levels of noise in the data, we then applied the Bandpass Frame Filtering method \citep[BFF;][]{Alzate2021}, with only the narrow band temporal filter (low-pass filter with cut-off at 18 hours) on the datacube obtained from the 6-hour averaged images.

The LASCO and Cor1 images in Figure~\ref{fig:overlays} were co-aligned and rotated to a solar-north up orientation based on image metadata. CATEcor data were oriented by comparing pre-eclipse sunspot images to observations from NASA's Solar Dynamics Observatory to determine the angle between the camera orientation and solar north. The images were matched in scale using the known pixel scale for the several instruments. However, with the CATEcor coronagraph adapter in use, it is not possible to determine the location of the Sun in the field of view, so images had to be hand-aligned based on the team's best estimates of the telescope pointing during the eclipse. Thus there is some uncertainty in the relative alignment of the images in the figure. (Modifications that will simplify the task of determining telescope pointing relative to the Sun with the occulter in place are already planned for an improved version of the CATEcor instrument currently in development.)

The top two panels of Figure~\ref{fig:overlays} show how the CATEcor observations match up against LASCO images; the full field of view is shown on the left, while on the right we cropped the image to fill the space covered by LASCO's occulter. Many of the features in the CATEcor images, particularly on the west side of the image, are artifacts. This is clear from the fact that they match poorly with features in LASCO, but can be confirmed by observing their motion in the animation accompanying Figure~\ref{fig:occulter_motion}. However, features on the north and east sides of the image
are strikingly similar to features visible in LASCO. These are features that also remain stationary even as the occulter moves, and thus are candidate coronal features.

The Cor1 images that directly overlap with CATEcor's field of view, seen in the bottom panels of the Figure, also bear some resemblance to features visible in the north and east in the CATEcor observations, although differences in perspective and the significant degradation of the Cor1 instrument make it difficult to conclusively determine if they are common features.

Nonetheless, given that these features are polarized to some extent, are fixed in the field of view even when the occulter moves, and match well with features seen in space-based coronagraphs, we conclude that CATEcor did successfully detect at least some of the K-corona.

\section{Discussion}
  \label{sec:discussion}
In building and deploying the initial CATEcor instrument at an expedition to the 14~October~2023 eclipse, we demonstrated a new type of coronagraph instrument.  The goals of the CATEcor project were threefold:
\begin{itemize}
    \item Use the annular eclipse as an opportunity to begin development of a simple coronagraph that could, eventually, be available to amateur observers who want to study the corona,
    \item Use the annular eclipse as both a chance to test Citizen CATE 2024 equipment in the field under realistic conditions and to carry out outreach regarding the project's plans for the 2024 total eclipse, and
    \item Use the annular eclipse to highlight the rich history of coronal physics in the United States in general and in the mountains of Colorado specifically.
\end{itemize}

In terms of these goals, CATEcor was an unmitigated success. Our eclipse expeditions showed the feasibility of observing the corona with relatively simple and inexpensive equipment, any time an annular or deep partial eclipse passes over a sufficiently high-enough vantage point. It also provided an opportunity for a real-world characterization of the CATEcor coronagraph \citep{DeForest2024}, and both the benefits and challenges of using it in the field.

The expeditions also provided some key lessons for future experiments. First, we did not initially anticipate the relative importance of the surface treatment of the occulters and support structure in reducing stray light. We learned that dark, matte surface treatments -- even with readily-available commercial spray paint -- are essential to minimizing stray light even in the favorable shaded-truss geometry. We anticipate that the new generation of novel ultra-black matte paints (e.g., the ``Black 4.0'' paint available from Stuart Semple's ``Culture Hustle'' art-supply company) will prove even more effective in future iterations of the instrument concept. Secondly, we learned that images from annular eclipses in particular and partial eclipses in general are more difficult to analyze than data from clear, coronal-condition sunny days: the crescent shape of the visible Sun during a partial eclipse introduced complex and highly structured patterns that proved difficult to interpret and to separate from the corona itself. This difficulty may outweigh the advantages of reduced sky brightness during a partial or annular eclipse; but more to the point, such eclipses are rare, and opportunities to use the instrument will be limited if it could only operate during eclipses. A future iteration of the instrument that can operate effectively under normal conditions would both eliminate the challenge introduced by these stray light artifacts and would be operable under a much wider range of conditions, and will be part of our focus in developing next-generation CATEcor instruments.

One other challenge proved a double-edged sword: while the occulter assembly's lack of stability helped us to interpret our observations, in spite of the complex stray light patterns in the data, lack of stability is nonetheless not desirable. Future iterations both need to improve and simplify the occulter alignment process, provide improved stability during observations, and could, in principle, support modifications that would improve tracking of the Sun's position during the observations. All of these improvements could simplify the image orientation and alignment process and, with the addition of multiple observations over time, might provide a path to much more robust separation stray light from signal with methods commonly used for more conventional coronagraphs, like the LASCO monthly minimum subtraction discussed in Section~\ref{sec:analysis}.

In spite of both these challenges and potential for future improvements, in the present, these expeditions indeed gave our team the chance to walk in the footsteps of giants, developing the kind of intuitive understanding that can only be gleaned from firsthand experience. As we did during our preliminary field test, Walter Orr Roberts struggled with the dusty conditions in the Colorado Rockies \citep{Bogdan2002}, a challenge magnified by his observatory's location near a working mine. More than 40 years before Menzel and Roberts set out to observe the corona at Climax, the pioneering optical physicist Robert W. Wood attempted unsuccessfully to detect the corona in daylight using polarimetric measurements \citep{Wood1900}. Wood himself was inspired by the eclipse of May 28, 1900, and found that his initial idea, of using narrowband observations of a specific spectral line, had been thoroughly (and similarly unsuccessfully) explored by George Hale.

Reading reports of these attempts illuminates these past struggles, but sharing in these struggles connects us to our scientific forebears much more deeply. Our experiments with CATEcor, and our expeditions to both a historic observing site in Colorado and to a popular, and near ideal, eclipse-track site in the mountains above Albuquerque, allowed us to immerse ourselves in the long scientific legacy of coronal observations both in the high country of Colorado and along eclipse tracks around the world. Likewise, the lessons we learned in following the footsteps of our predecessors, coupled with technological progress that has enormously simplified fabrication, deployment, and operation of astronomical instrumentation, allowed us to demonstrate the feasibility of a low-barrier-to-entry shaded-truss coronagraph.

Continued investment in such accessible research, and in research at solar eclipses in particular, is critical to fostering this deep understanding of solar physics' roots, providing outreach opportunities that inspire new researchers to enter the field, and kindle innovative instruments and approaches -- which are, in turn, needed to gather novel data and answer new questions.

\acknowledgments
This paper is dedicated to the memory of Prof. Jay M. Pasachoff of Williams College, who served as mentor to two of us (Seaton '01 \& Tosolini '23), and whose dedication to eclipse science and outreach inspired all of us.

We are grateful to many people and organizations who helped to make this observation possible. Citizen CATE 2024 is jointly funded by the National Science Foundation (Grant Nos. 2231658, 2308305, \& 2308306), NASA (Grant Nos. 80NSSC21K0798 \& 80NSSC23K0946), and Southwest Research Institute.  
The CATEcor telescope occulter was developed under internal funding by Southwest Research Institute 
using open-source tools including FreeCAD
(https://freecad.org), and the expedition was funded by the Citizen CATE 2024 project and by SwRI.  

The Colorado expedition was enabled by free access to the White River National Forest, access to which remains one of Colorado's greatest assets. Daniel Zietlow assisted greatly with eclipse-day logistics, documentation, and outreach activities. David Elmore provided both historical documents about, and his own first-hand experience of, the Climax Observatory. Scott McIntosh was enormously helpful in our attempts to facilitate access to the Climax site.

The New Mexico expedition was enabled by kind interactions with the Cibola National Forest and National Grasslands office, and specifically District Ranger Crystal Powell, of the Sandia Ranger District, who clarified access conditions and helped us with site selection.  We were pleased to collaborate both with Windfall Films of London (UK), specifically Joby Lubman and his crew, and the Logan School for Creative 
Learning of Denver (CO), specifically Catherine Peterson and her team, in engaging  students and the general public in the event.  We are grateful to the PUNCH mission Outreach team, specifically Cherilynn Morrow, for making available several important learning tools including the ``three-hole PUNCH pinhole projector'' that we 
distributed in large numbers.

\begin{authorcontribution}
DB: conceived the CATEcor expeditions and led development of the project and this writeup, led the Loveland Pass dry run and expedition, and contributed significantly to the design of the CATEcor adapter;

AC: leads the Citizen CATE 2024 project, helped with the development and logistics for the expedition, and contributed significantly to the design of the CATEcor adapter;

SK: managed preparation for the expeditions, manages deployment logistics for the Citizen CATE 2024 project, led data acquisition at the Sandia Crest site, and led shakeout and testing of the CATEcor adapter;

CD: designed, produced, assembled, and tested the CATEcor adapter; led the Sandia Crest expedition, and contributed significantly to writeup;

AT: led eclipse-data data acquisition at the Loveland Pass site;

NA: Processed STEREO Cor1 data and contributed to the writeup;

All other authors played a significant role in the expeditions, preparation, data acquisition, data analysis, and writeup.
\end{authorcontribution}

\begin{fundinginformation}
CATEcor development and deployment, including the eclipse expeditions, was funded in part by the National Science Foundation under Grants 2231658 and 2308305, and in part by Southwest Research Institute under internal funding.  Funding for Citizen CATE 2024 was provided by grants from the NSF (awards 2231658, 2308305, \& 2308306) and NASA (grants  80NSSC21K0798 \& 80NSSC23K0946).
\end{fundinginformation}

\begin{dataavailability}
Unprocessed CATEcor images are available in TIFF format on request.
\end{dataavailability}

\begin{materialsavailability}
All Citizen CATE 2024 equipment is available off the shelf, and is described in \citet{Patel2023}. A future paper, following the 2024 total eclipse will describe performance and capabilities of the telescope/camera setup in detail. The CATEcor coronagraph, including references to detailed 3D print designs suitable for download, is described in detail in \citet{DeForest2024}.
\end{materialsavailability}

\begin{codeavailability}
Both the SWAP filter and NRGF are publicly available via SolarSoft IDL and/or Python packages. Additional custom code used to process CATEcor data and produce these images is available on request to the authors.
\end{codeavailability}

\begin{ethics}
\begin{conflict}
The authors declare that they have no conflicts of interest or competing interests.
\end{conflict}
\end{ethics}


\end{document}